\documentclass[
reprint,  floatfix,
amsmath,amssymb,aps,showpacs,superscriptaddress,dvipdfm   % 
nofootinbib, showkeys
]{revtex4-1}
\usepackage{amssymb}
\usepackage{graphicx}
\usepackage{booktabs}
\usepackage{dcolumn}% Align table columns on decimal point
\usepackage{bm}% bold math
\usepackage{mathrsfs}
\usepackage{mathtools}
\usepackage{amsmath}
\usepackage{amssymb}
\usepackage{epstopdf}
\usepackage{units}
\usepackage{mathtools}
\usepackage{ragged2e}
\usepackage[mathlines]{lineno}% Enable numbering of text and display math
\usepackage{dsfont}
\usepackage{upgreek}
\usepackage{subfig}
\begin{document}
\preprint{APS/123-QED}

\title{Experimental metrics for detection of detailed balance violation}

\author{Juan Pablo Gonzalez}
\affiliation{Duke University, Department of Physics, Box 90305
Durham, NC 27708-0305}

\author{John C. Neu}
\affiliation{University of California, Berkeley, Department of Mathematics, Berkeley, CA 94720-3840}

\author{Stephen W. Teitsworth}
\affiliation{Duke University, Department of Physics, Box 90305
Durham, NC 27708-0305}

\date{\today}

\begin{abstract}
We report on the measurement of detailed balance violation in a coupled, noise-driven linear electronic circuit consisting of two nominally identical RC elements that are coupled via a variable capacitance. The state variables are the time-dependent voltages across each of the two primary capacitors, and the system is driven by independent noise sources in series with each of the resistances. From the recorded time histories of these two voltages, we quantify violations of detailed balance by three methods: 1) explicit construction of the probability current 
density, 2) by constructing the time-dependent stochastic area, and 3) by constructing 
statistical fluctuation loops.  In comparing the three methods, we find 
that the stochastic area is relatively simple to implement, computationally 
inexpensive, and provides a highly sensitive means for detecting 
violations of detailed balance.

\end{abstract}

\pacs{}
\keywords{noise, detailed balance, far-from-equilibrium}
\maketitle
%\tableofcontents

%%%%%%%%%%%%%%%%%%%%%%%%%%%%%%%%%%%%%%%%%%%%%%%%%%%%%%%%%%%%%%%%%%%%%%%%%%%%
%%%%%%%%%%%%%%%%%%%%%%%%%%%%%%%%%%%%%%%%%%%%%%%%%%%%%%%%%%%%%%%%%%%%%%%%%%%%
%%%%%%%%%%%%%%%%%%%%%%%%%%%%%%%%%%%%%%%%%%%%%%%%%%%%%%%%%%%%%%%%%%%%%%%%%%%%
%%%%%%%%%%%%%%%%%%%%%SECTION ONE%%%%%%%%%%%%%%%%%%%%%%%%%%%%%%%%%%%%%%%%%%%%
%%%%%%%%%%%%%%%%%%%%%%%%%%%%%%%%%%%%%%%%%%%%%%%%%%%%%%%%%%%%%%%%%%%%%%%%%%%%
%%%%%%%%%%%%%%%%%%%%%%%%%%%%%%%%%%%%%%%%%%%%%%%%%%%%%%%%%%%%%%%%%%%%%%%%%%%%
\section{Introduction}
\label{sec:intro}

Detailed balance violation is an essential feature of many non-equilibrium systems.  In the context of noise-driven dynamical systems, detailed balance violation generally implies a non-vanishing steady state probability current in 
the system phase space \cite{Tolman1938, Gardiner_2009, Zia_Schmittmann_2007, vanKampen_2007}. Additionally, violations of detailed balance often indicate that the system is ``open," i.e., subject to external driving forces which induce energy transfer through it.  Examples abound in diverse fields such as climate dynamics 
\cite{Weiss_Tellus_2003, Weiss_PRE_2007, Penland_JClimate_1995, Newman_2017}, active biological systems 
\cite{Battle_Science_2016, Gladrow_2016, Gnesotto_2018, Mura_2018}, electronic transport systems 
\cite{Bomze_PRL_2012, Ciliberto_PRL_2013, Ciliberto_JSM_2013b, Chiang_2017b}, micromechanical oscillators 
\cite{Chan_PRL_2008, Chan_PRE_2008}, and microscopic heat engines \cite{Martinez_2017}. 
The fluctuation statistics of voting models \cite{Mellor_EPL_2016} and financial markets 
\cite{Fiebig_2015} also display behavior that is analogous to detailed balance violation observed in the aforementioned physical systems.  The common behaviors observed in these systems motivate the development of widely applicable metrics that can quantify the level of detailed balance violation in far-from-equilibrium systems.

The construction of probability current from numerical or experimental data is a classic indicator of detailed balance violation.  Due to 
conservation of probability, the steady probability current 
is divergence-free, so it typically has a 
circulating structure.  This tendency has been confirmed in numerous 
theoretical studies \cite{Filliger_2007, Zia_Schmittmann_2007, Mellor_EPL_2016, Gladrow_2016, Ghanta2017}.   Experiments on such systems as actively beating flagella and thermally driven electrical circuits have directly measured circulating probability currents \cite{Battle_Science_2016, Chiang_2017a, 
Chiang_2017b}.  These experiments can be challenging because they require a great deal of data in order to define the vector field on a fine enough grid and with sufficient 
number of data points for each grid location.  

\begin{figure*}[t]
\includegraphics[width=1.0\textwidth]{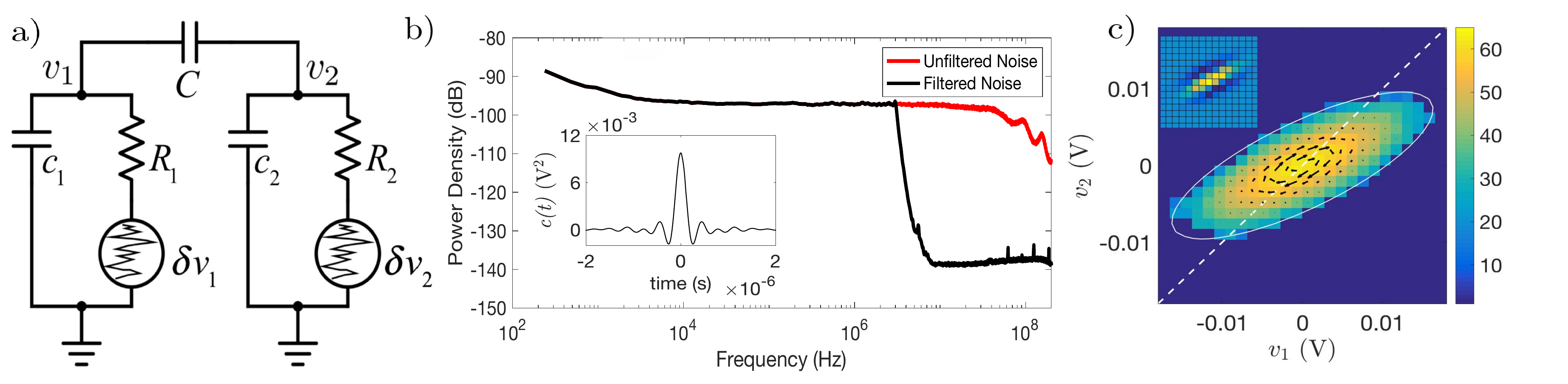}
\caption{
(a) Experimental circuit diagram.  (b) Power spectrum of noise generator before and after bandpass filtering. Inset shows the measured temporal autocorrelation function $c(t)$ of filtered noise.  (c) Experimentally measured steady state probability density and probability current for asymmetric applied noise intensities, i.e., $s_{1}^2 = 8.53\times 10^{-10}$ V$^{2}\cdot$s and  $s_2^2 = 5.30\times 10^{-11}$ V$^{2}$s.  }
\label{fig:CircuitDiagram}
\end{figure*}

In this paper, we demonstrate alternative experimental metrics of detailed balance violation which are easier to implement and more sensitive.
The experiments are conducted on an electrical circuit consisting of two nominally identical $RC$ circuit elements that are driven by independent noise sources and capacitively coupled to one 
another.  When the noise sources have unequal intensities we  
observe violations of detailed balance through direct measurements of 
circulating probability currents in the system phase space.  Alternatively, we use the experimental data to construct the time-dependent \textit{stochastic area} recently introduced by Ghanta \textit{et al.} \cite{Ghanta2017}.  The construction of stochastic area is much easier and, as a metric of departures from detailed balance, much more sensitive.  The sensitivity derives from its \textit{global} character:  it uses \textit{all}
the data from a long running experiment.  This contrasts with the probability current density vector field in a certain pixel of phase space which utilizes only data corresponding to the portion of the system trajectory in the given grid box.

We demonstrate that experimental ensemble-averaged fluctuation loops \cite{Ghanta2017} can be constructed using an equivalent amount 
of data as used for probability current.  Such loops are intimately 
connected with the geometric theory of large deviations 
 \cite{WentzellFreidlin_2012, Luchinsky_RPP_1998, Heymann_2015, 
Neu_Geometry_2018}. The measured loops allow one to 
quantitatively visualize that nature of fluctuations from the highly 
probable stable fixed point to rarely occurring remote states and the 
relaxation back to the fixed point, a dynamics that cannot be inferred from 
a plot of probability current alone!

The structure of the paper is as follows.  Section II describes the design of the experiment, its physical construction, and the procedures to measure steady probability density and current.  It finishes with an overview of the circuit model and its predictions of the steady state probability density and current.   Section III presents measurements of the stochastic area for different noise combinations which either violate or satisfy detailed balance.   We discuss the merits of stochastic area as an experimental metric of detailed balance violation, relative to probability current, as well as the importance of choosing appropriate sampling rates.  Section IV addresses the dependence of the fluctuation statistics upon system parameters.  Specifically, the model in Sec. II clearly predicts how the probability density and time rate of change of stochastic area vary with the coupling capacitance; experimental data confirm these predictions.  In particular, we find that it is possible to ``tune" the coupling capacitance so that the time rate of change of stochastic area is maximized.  Section V presents experimental constructions of fluctuation loops and their significance.  Section VI discusses the connections between the stochastic area and other related metrics for characterizing detailed balance violation in nonequilibrium systems \cite{Mura_2018, Ciliberto_PRL_2013, Ciliberto_JSM_2013b}.  A concluding section includes a brief discussion of the connection between the stochastic area and seminal work of Onsager on thermodynamic correlation functions \cite{Onsager_1931}.

%%%%%%%%%%%%%%%%%%%%%%%%%%%%%%%%%%%%%%%%%%%%%%%%%%%%%%%%%%%%%%%%%%%%%%%%%%%%
%%%%%%%%%%%%%%%%%%%%%%%%%%%%%%%%%%%%%%%%%%%%%%%%%%%%%%%%%%%%%%%%%%%%%%%%%%%%
%%%%%%%%%%%%%%%%%%%%%%%%%%%%%%%%%%%%%%%%%%%%%%%%%%%%%%%%%%%%%%%%%%%%%%%%%%%%
%%%%%%%%%%%%%%%%%%%%%SECTION TWO%%%%%%%%%%%%%%%%%%%%%%%%%%%%%%%%%%%%%%%%%%%%
%%%%%%%%%%%%%%%%%%%%%%%%%%%%%%%%%%%%%%%%%%%%%%%%%%%%%%%%%%%%%%%%%%%%%%%%%%%%
%%%%%%%%%%%%%%%%%%%%%%%%%%%%%%%%%%%%%%%%%%%%%%%%%%%%%%%%%%%%%%%%%%%%%%%%%%%%
\section{Experimental Setup and Dynamical Circuit Model}
\label{sec:circuit}

The experimental system is a linear electrical circuit comprised of two nominally identical $RC$ sections that are capacitively coupled to one another.  Each $RC$ section is driven by an independent noise source, cf. Fig. \ref{fig:CircuitDiagram}(a) for the schematic.  The coupled $RC$ network is built on a circuit breadboard and secured inside a metal box fitted with coaxial connections to avoid external interference.  The resistances $R_1$ and $R_2$ are metal film type, and the capacitances $C, c_1$, and $c_2$ are ceramic disc capacitors. Parameter values used in the measurements reported here are $R_1 \approx R_2 := R = 1.20 \text{ k}\Omega$, $c_1 \approx c_2 = c := 33.1$ nF, and coupling capacitances in range $C =$ 100 pF - 880 nF. All nominally identical components are verified to have parameter values within 1 \% of each other.  To measure the voltage variables $v_1$ and $v_2$, we use a Picoscope 2406B, a compact, computer-controlled oscilloscope that serves as a multichannel analog-to-digital converter. It reads the continuous time voltage signals $v_1(t)$ and $v_2(t)$ at a regular sampling interval $\tau$, thereby yielding discrete voltage readings $v_1[t]$ and $v_2[t]$, with 8-bit vertical resolution and a sampling rate up to 250 MS/s. Acquired data from the picoscope are analyzed using MATLAB.  

Injected noise signals $\delta v_1$ and $\delta v_2$ are created using a dual channel Tektronix AFG3252 arbitrary function generator with 240 MHz bandwidth.  The effective amplitudes can be varied independently for each injected noise term.  To avoid high frequency parasitic effects in the circuit, the function generator output signals are each passed through nominally identical low pass filters (Mini-Circuits Model: SLP-2.5+) with cut-off frequency $f_c =$ 2.5 MHz \footnote{If the low pass filters are not used, we find that the high frequency components of the noise couple to the parasitic inductances associated with the resistors (typically of order a few $\mu$H) and this, in turn, gives rise to a more complex stochastic dynamics of the circuit.  This can be understood by recognizing that each additional inductance adds an extra phase space dimension to the corresponding circuit model.}. Fig. \ref{fig:CircuitDiagram}(b) shows the measured power spectrum of the noise generator output both before and after filtering.  The inset shows the measured autocorrelation of the filtered noise signal.  It is clearly symmetric under time inversion and the central peakwidth provides a measure of the correlation time $t^* \approx 400$ ns. Provided that the correlation time $t^*$ is much smaller than the deterministic relaxation time $Rc \approx$ 40 $\mu$s, the injected experimental noises are well-described as delta-correlated white noises in the circuit model presented below.
 
By placing the noise sources in series with the resistors, we have in mind the natural thermal (Johnson-Nyquist) noises.  In our experiment, intrinsic thermal noises are negligibly small relative to the added noises.  In contrast, recent experiments by Chiang et al. studied stochastic gyrating dynamics in a similar circuit system using natural thermal noises \cite{Chiang_2017a, Chiang_2017b}.  By placing one of the $RC$ elements in a cryogenic environment  and employing relatively large resistances on the order of $10^6 \text{ } \Omega$, the thermal noise voltages are large enough to allow measurement of steady probability currents and detection of detailed balance violation.  Relative to the experiments reported in this paper, using larger resistances implies longer circuit relaxation times $Rc$, and proportionately longer times to collect sufficient data.

We now describe the processing of voltage time series which generates measured approximations to the steady probability density and current in the $v_1 -v_2$ plane.  First, identify a region in this plane which contains almost all of the sample points, but whose dimensions are still comparable to the standard deviations of $v_1$ and $v_2$.  This region is divided into square ``pixels" with dimensions small compared to the aforementioned standard deviations.  Next, construct the \textit{histogram} which records the number of data points in each pixel.  The approximate probability density is the \textit{fraction} of data points in each pixel.  Figure \ref{fig:CircuitDiagram}(c) shows a typical measured probability distribution in the $v_1-v_1$ phase plane which results from the noise strength in $RC$ element 1 exceeding that of 2.  The color scale is lograthmically scaled since the probability distribution has a Gaussian profile.

The probability current is approximated by a \textit{vector-valued} histogram:  From the original voltage time series $v_1[t]$ and $v_2[t]$,
construct the time series of displacement vectors
\begin{equation}
\begin{pmatrix}
\Delta v_1[t] \\
\Delta v_2[t]
\end{pmatrix}
= \begin{pmatrix}
v_1[t + \tau] - v_1[t] \\
v_2[t + \tau] - v_2[t]
\end{pmatrix}
.
\end{equation}
Assign each displacement vector to the pixel in which it occurs.  After a sufficiently long run time $T$, the probability current in a given pixel is approximated by the vector sum of displacements in that pixel, divided by $T$, and the area of the pixel.  This is how the arrows representing probability current in Fig. \ref{fig:CircuitDiagram}(c) are generated.  Although this construction is simple and intuitive, how do we really know that it generates the probability current?  The answer comes from an analysis of the stochastic ODE which models fluctuation statistics in the $v_1 -v_2$ plane.  

Using the Kirchhoff laws, the dynamical circuit model for voltages $v_1(t)$ and $v_2(t)$ is
\begin{equation}
\label{eq:Kirchhoff}
R
\begin{pmatrix}
C + c & -C \\
-C & C + c
\end{pmatrix}
\begin{pmatrix}
\dot{v}_1 \\
\dot{v}_2
\end{pmatrix} = -
\begin{pmatrix}
v_1 \\
v_2
\end{pmatrix}
+
\begin{pmatrix}
s_1(t) \\
s_2(t)
\end{pmatrix}
.
\end{equation}
Here, $s_1(t)$ and $s_2(t)$ are the injected noise signals in series with each resistor, and corresponding to $\delta v_1$ and $\delta v_2$, resp., in the experimental schematic, cf. Fig. \ref{fig:CircuitDiagram}(a).  Since the noise correlation time is much shorter than the $Rc$ relaxation time, but not so short as to induce high frequency parasitics, for modeling purposes, we can express the actual noises $s_1(t)$ and $s_2(t)$ in (\ref{eq:Kirchhoff}) as
\begin{equation}
\label{eq:noises}
s_i(t) \approx s_{i}w_{i}(t), i = 1,2,
\end{equation}
where $w_{1}(t)$ and $w_{2}(t)$ are independent unit white noises, i.e., $\langle w_{i}(t)w_{j}(t')\rangle=\delta_{ij} \delta(t-t')$. The coefficients $s_1$ and $s_2$ are \textit{noise amplitudes}.  Note that detailed balance is broken by taking the noise amplitudes to be different from one another.  The assignment of effective noise amplitudes to the filtered experimental noise signals is described in Appendix A.  

The circuit model (\ref{eq:Kirchhoff}) is now expressed as
\begin{equation}
\label{eq:CircuitModel}
R
\begin{pmatrix}
C + c & -C \\
-C & C + c
\end{pmatrix}
\begin{pmatrix}
\dot{v}_1 \\
\dot{v}_2
\end{pmatrix} = -
\begin{pmatrix}
v_1 \\
v_2
\end{pmatrix}
+
\begin{pmatrix}
s_1 & 0 \\
0 & s_2
\end{pmatrix}
\begin{pmatrix}
w_1(t) \\
w_2(t)
\end{pmatrix}
,
\end{equation}
which is equivalent to a stochastic differential equation of form
\begin{equation}
\dot{\mathbf{v}} = L \mathbf{v} + \sigma \mathbf{w},
  \label{eq:SDE}
  \end{equation}
where $\mathbf{v} :=
\begin{pmatrix}
v_1 \\
v_2
\end{pmatrix}$ is the state vector and $\mathbf{w} :=
\begin{pmatrix}
w_1 \\
w_2
\end{pmatrix}$ denotes the vector of independent unit white noises.  Comparing (\ref{eq:CircuitModel}) and (\ref{eq:SDE}), we identify the \textit{dynamical tensor} $L$,
\begin{equation}
L^{-1} = -R
\begin{pmatrix}
C + c & -C \\
-C & C + c
\end{pmatrix},
\end{equation}
and state-independent \textit{noise tensor} $\sigma$,
\begin{equation}
\sigma = -L
\begin{pmatrix}
s_1 & 0 \\
0 & s_2
\end{pmatrix}.
\end{equation}

Due to the linearity of the stochastic dynamics, the steady state probability density is a Gaussian proportional to
\begin{equation}
\label{eq:ProbDensity}
\rho(\mathbf{v}) \propto \exp (-\frac{1}{2} \mathbf{v}^T M^{-1} \mathbf{v}),
\end{equation}
where $M$ denotes the \textit{second moment tensor} with components $M_{ij} = \langle v_i v_j \rangle$.  The second moment tensor is determined by a \textit{fluctuation-dissipation} relation, which implies that $LM + D$ is antisymmetric \cite{Weiss_Tellus_2003, Ghanta2017}.  Here $D$ is the \textit{diffusion tensor}, 
\begin{equation}
D := \sigma \sigma^T = LSL, \: \: S := \begin{pmatrix}
s_1^2 & 0 \\
0 & s_2^2
\end{pmatrix}.
\end{equation}
The fluctuation-dissipation relation amounts to linear inhomogeneous equations for the components of $M$.  Physically, they express the balance between flow towards the origin, embodied by the dynamical tensor $L$ (the dissipation) and spreading (the fluctuation) embodied by the diffusion tensor $D$.  These equations determine the components of $M$ as functions of the circuit parameters and noise amplitudes.  

The general expression for probability current density is
\begin{equation}
\label{eq:ProbCurrent1}
\mathbf{j} = L \mathbf{v} \rho - D \nabla_{\mathbf{v}} \rho.
\end{equation}
For the stationary probability as in (\ref{eq:ProbDensity}), we have
\begin{equation}
\nabla \rho = -M^{-1} \mathbf{v} \rho \nonumber,
\end{equation}
and then the stationary probability current is 
\begin{equation}
\label{eq:StaProbCurrent}
\mathbf{j} = (L + DM^{-1})\mathbf{v} \rho = (LM + D)M^{-1}\mathbf{v} \rho.
\end{equation}
Due to antisymmetry of $LM + D$ we can, in two dimensions, write the simple form 
\begin{equation}
LM + D = \begin{pmatrix} 0 & -\Omega \\ \Omega & 0 \end{pmatrix},
\end{equation}
where $\Omega$ denotes the stochastic vorticity $\Omega = -(LM + D)_{12}$ \cite{Ghanta2017}.
This allows us to express the stationary probability current density as 
\begin{equation}
\label{eq:ProbCurrent2}
\mathbf{j} = \Omega \begin{pmatrix} 0 & -1 \\ 1 & 0 \end{pmatrix} M^{-1} \mathbf{v} \rho.
\end{equation}

We outline the mathematics behind the construction of probability current as a "vector-valued" histogram.   Let $\delta R$ be a fixed small region in the $v_1 - v_2$ plane, such as one of the pixels of the histogram.  A stochastic trajectory makes several intermittent transits of the region $\delta R$ in the time interval $0 < t < T$.  For each transit, record the change $\Delta \mathbf{v}$ in $\mathbf{v}$ between entry and departure.  It can be shown \cite{Neu_2018} that the ensemble-averaged sum of these  $\Delta \mathbf{v}$ divided by $T$ equals the integral of probability current over $\delta R$, i.e., $\int_{\delta R} \mathbf{j} \: d^2v$.

%%%%%%%%%%%%%%%%%%%%%%%%%%%%%%%%%%%%%%%%%%%%%%%%%%%%%%%%%%%%%%%%%%%%%%%%%%%%
%%%%%%%%%%%%%%%%%%%%%%%%%%%%%%%%%%%%%%%%%%%%%%%%%%%%%%%%%%%%%%%%%%%%%%%%%%%%
%%%%%%%%%%%%%%%%%%%%%%%%%%%%%%%%%%%%%%%%%%%%%%%%%%%%%%%%%%%%%%%%%%%%%%%%%%%%
%%%%%%%%%%%%%%%%%%%%%SECTION THREE%%%%%%%%%%%%%%%%%%%%%%%%%%%%%%%%%%%%%%%%%%
%%%%%%%%%%%%%%%%%%%%%%%%%%%%%%%%%%%%%%%%%%%%%%%%%%%%%%%%%%%%%%%%%%%%%%%%%%%%
%%%%%%%%%%%%%%%%%%%%%%%%%%%%%%%%%%%%%%%%%%%%%%%%%%%%%%%%%%%%%%%%%%%%%%%%%%%%
\section{Experimental determination of stochastic area}

For the circuit system (\ref{eq:SDE}), the stochastic area is defined by the line integral \begin{equation}
        A(t) = \frac{1}{2} \int_{C(t)}^{ } (v_1dv_2 - v_2dv_1),
        \label{eq:AT_theory}
    \end{equation}
where $C(t)$ denotes a specific stochastic trajectory of the system over the time range from $0$ to $t > 0$.  Geometrically, the stochastic area is simply the area swept out by the trajectory in the $v_1-v_2$ plane over the time interval $(0,t)$. In Ghanta \cite{Ghanta2017}, it is shown that its stationary ensemble average rate of change is precisely the prefactor $\Omega$ of the probability current in (\ref{eq:ProbCurrent2}),
\begin{equation}
\label{eq:Adot}
\langle \dot{A} \rangle = \Omega=-(LM + D)_{12}.
\end{equation}
Since $\mathbf{j} \equiv 0$ iff $\Omega = 0$, the stochastic area is a clear detector of detailed balance violation.  In contrast to the probability density vector field, stochastic area is a \textit{global} property of voltage fluctuation statistics.

Given experimentally recorded voltage time series $v_1[t]$ and $v_2[t]$ with sampling interval $\tau$, the natural finite difference approximation to the time rate of change
\begin{equation}
\dot{A} (t) = \frac{1}{2}(v_1 \dot{v}_2 - v_2 \dot{v}_1) (t) \nonumber
\end{equation}
of stochastic area is
\begin{equation}
\label{eq:DA_exp}
(DA)[t] := \frac{1}{2 \tau} \{ v_1[t] v_2[t + \tau] - v_1[t + \tau] v_2[t] \}.
\end{equation}
Hence, the discrete approximation to the stochastic area at time $t = N \tau$ is
\begin{equation}
\label{eq:AT_exp}
A[t] = \frac{1}{2} \sum_{k = 0}^{N-1} (DA)[k \tau].
\end{equation}
Figure 2 is a visualization of the discrete area, which results from linear interpolation between successive measurement points $(v_1[t], v_2[t])$.

An analysis of the stochastic ODE (\ref{eq:CircuitModel}) leads to an explicit expression for the ensemble average of $DA$ in (\ref{eq:DA_exp}),
\begin{equation}
\label{eq:DAtheory}
\langle DA \rangle = \frac{1}{2 \tau} (Me^{\tau L} - e^{\tau L}M)_{12}.
\end{equation}
See Appendix B for the derivation.  In the limit of sampling interval $\tau$ much shorter than the relaxation time $Rc$, $\langle DA \rangle$ converges to the theoretical prediction (\ref{eq:Adot}), that is,
\begin{equation} 
\langle DA \rangle \rightarrow -(LM - ML)_{12} = -(LM + D)_{12} = \Omega.
\end{equation}
    \begin{figure}[h]
        \includegraphics[scale=0.25]{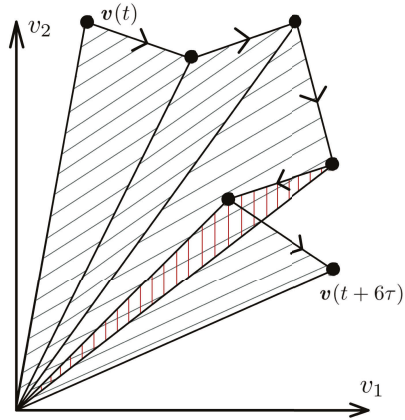}
        \caption{
        Schematic illustration of the geometric construction of experimental stochastic area $A(t)$ with sampling interval $\tau$.}
        \label{fig:DiscreteArea}
    \end{figure}
    
\begin{figure}[h]
\includegraphics[width=1.0\columnwidth]{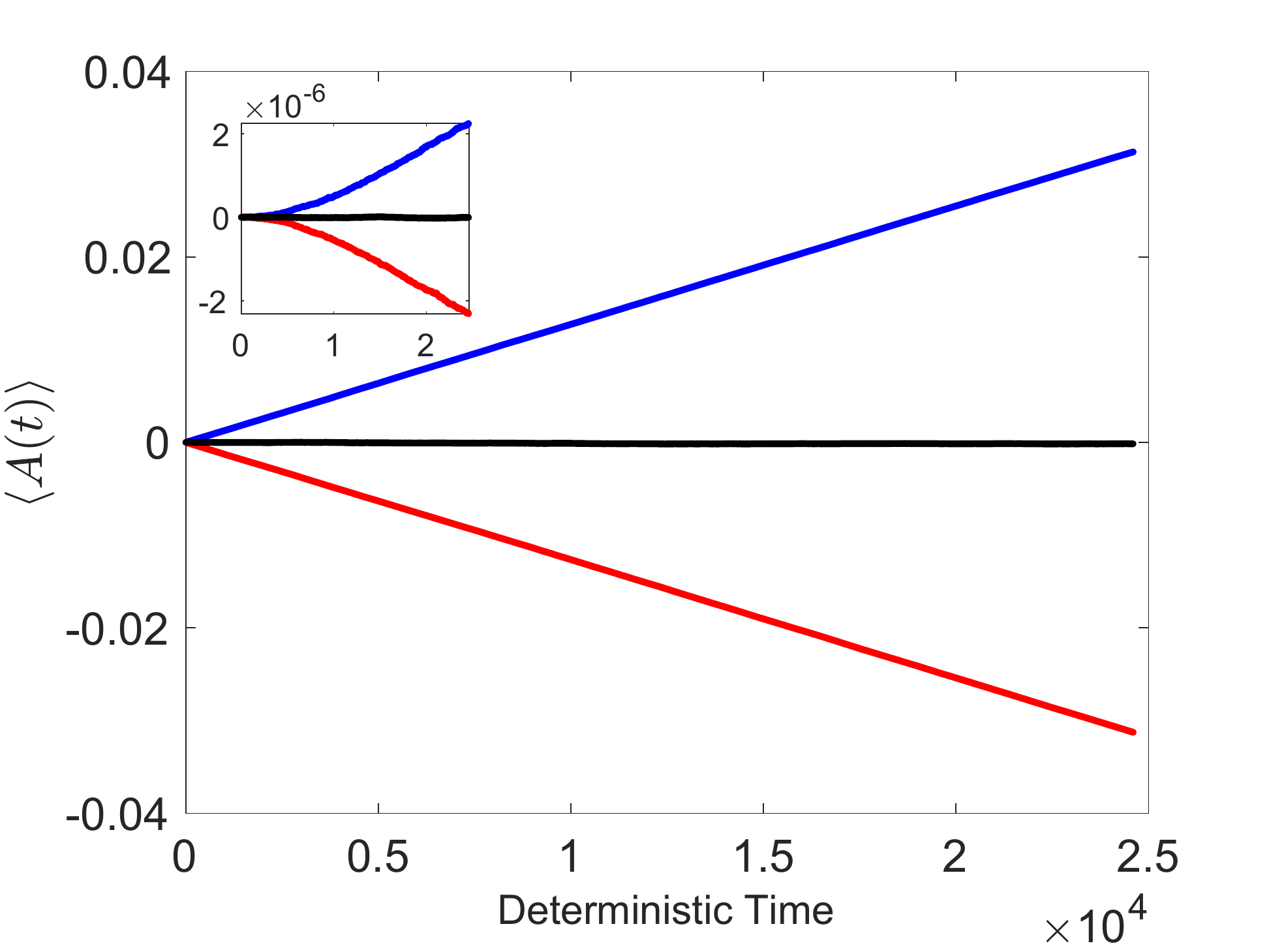}
\caption{(Color Online) Time-dependent ensemble-averaged stochastic area showing linear dependence on time for three distinct pairs of noise intensities at long times: $s_{1}^2 < s_{2}^2$ for upward sloping curve (blue), $s_{1}^2 > s_{2}^2$ for the downward sloping curve (red), and $s_{1}^2 = s_{2}^2$ for the horizontal curve (black).  The inset shows \textit{quadratic} time dependence of the ensemble-averaged stochastic area at relatively short times for the same pairs of noise intensities.}
\label{fig:AreaTime}
\end{figure}

Figure \ref{fig:AreaTime} shows typical experimental results for the time dependence of $\langle A(t) \rangle$ extracted from the recorded time series according to (\ref{eq:DA_exp}) and (\ref{eq:AT_exp}). In this case, the sampling interval $\tau$ is $500ns << Rc \simeq 40 \mu s$ and each curve in the main graph of Figure \ref{fig:AreaTime} is the result of averaging over 25 trials.  The values of the applied noise intensities are, respectively, $s_1^2 = 8.53 \times 10^{-10}$ $\mathrm{V^2 s}$ and $s_2^2 = 5.30 \times 10^{-11}$ $\mathrm{V^2 s}$ for the positive slope curve, $s_1^2 = 5.30 \times 10^{-11}$ $\mathrm{V^2 s}$ and $s_2^2 = 8.53 \times 10^{-10}$ $\mathrm{V^2 s}$ for the negative slope curve, and $s_1^2 = 4.82 \times 10^{-10}$ $\mathrm{V^2 s}$ and $s_2^2 = 4.86 \times 10^{-10}$ $\mathrm{V^2 s}$ for the nominally horizontal curve. The graph with $s_1 > s_2$ exhibits \textit{positive} slope, consistent with the clockwise circulation of the probability current in Fig. \ref{fig:CircuitDiagram}(b). For $s_1 < s_2$, the graph shows \textit{negative} slope of equivalent magnitude and consistent with the counterclockwise probability current. Approximately equal noise amplitudes $s_1 \approx s_2$ is close to detailed balance and yields a horizontal slope \footnote{The small difference in $s_i$ values stems from a slight difference in output of the two noise channels in the function generator for nominally identical output settings.  Interestingly, if one increases the number of ensemble averages is it possible to detect the small violation of the detailed balance in the area curve.  See also the discussion in the Appendix on determination of experimental $s_i$ values.}.  The overall length of the time series is of order 1 sec, much longer than the relaxation time $Rc \simeq 40$ $\mu$s.  For all three curves, the measured area curves are in close agreement with the theoretical prediction based on (\ref{eq:Adot}).  It is remarkable that the experimental curves are so close to predicted behavior and with a relatively small number of averages;  this attests to the robustness of the stochastic area as an experimental tool and suggests that it may be usefully applied to other detailed balance violating systems.  
    
The inset of Fig. \ref{fig:AreaTime} shows the experimental behavior of ensemble-averaged stochastic area at short times, i.e., times smaller than the deterministic relaxation time.  The transition from quadratic to linear behavior as time increases is evident and this behavior is consistent with earlier theoretical predictions \cite{Ghanta2017}.  For this data we must use a smaller sampling time $\tau = 50$ ns and average over 1000 trials.  To accurately capture short time behavior which is more sensitive to the injected noise, we find that it is typically necessary to average over a much larger set of trials than for the long time behavior.  
    
\begin{figure}[h]
\includegraphics[scale=0.45]{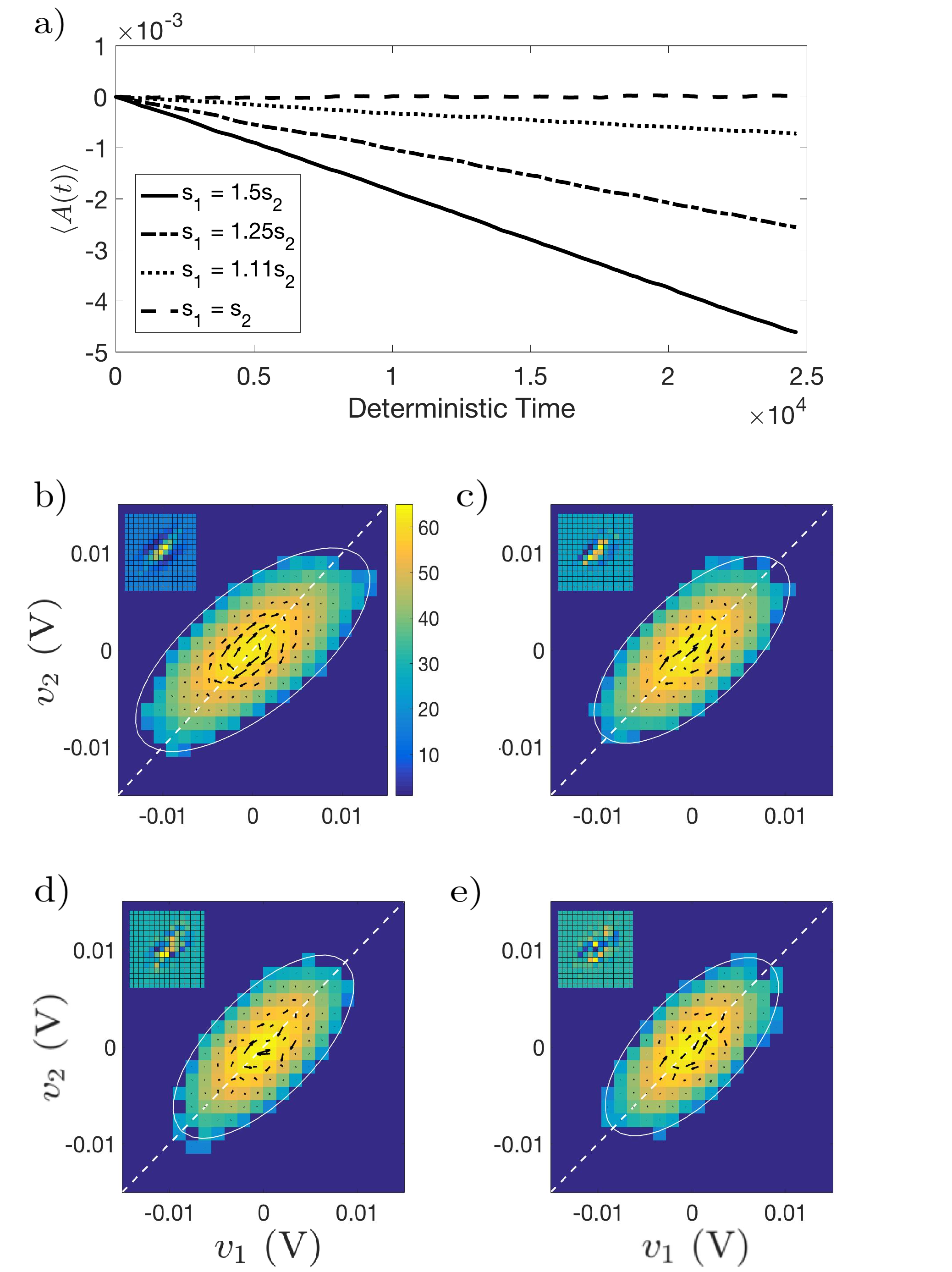}
\caption{(a) Time-dependent ensemble-averaged stochastic area for four distinct pairs of noise intensities at long times.  (b) - (e) Corresponding experimental probability densities and currents, where insets show the measured axial components of the curl of the probability current.}
\label{fig:NoiseDiff}
\end{figure}

Figure \ref{fig:NoiseDiff} compares the detectability of detailed balance violation using the stochastic area versus probability current for successively smaller values in the difference of the noise intensities (i.e., $s_1^2 - s_2^2$). For a sufficiently large difference in $s_i$ values (see, e.g., the stochastic area curve with $s_1 = 1.5 s_2$ in Fig. \ref{fig:NoiseDiff}(a) and corresponding probability current density of Fig.\ref{fig:NoiseDiff}(b)) the violation of detailed balance is clear in both sets of data. However, as the difference is reduced, the detection of detailed balance violation becomes much more challenging when based on probability current density measurement alone.  This is illustrated by comparing the stochastic area curve with $s_1 = 1.11 s_2$ in Fig. \ref{fig:NoiseDiff}(a) with the corresponding probability density current in Fig. \ref{fig:NoiseDiff}(d). The area curve shows a clear positive slope (with only 25 averages!) while the probability current density and curl are essentially indistinguishable from the detailed balance case shown in Fig. \ref{fig:NoiseDiff}(e).  In principle, the  probability current histograms can be improved by averaging over more trials, but the effort becomes prohibitive as the mesh of pixels is progressively refined.

%%%%%%%%%%%%%%%%%%%%%%%%%%%%%%%%%%%%%%%%%%%%%%%%%%%%%%%%%%%%%%%%%%%%%%%%%%%%
%%%%%%%%%%%%%%%%%%%%%%%%%%%%%%%%%%%%%%%%%%%%%%%%%%%%%%%%%%%%%%%%%%%%%%%%%%%%
%%%%%%%%%%%%%%%%%%%%%%%%%%%%%%%%%%%%%%%%%%%%%%%%%%%%%%%%%%%%%%%%%%%%%%%%%%%%
%%%%%%%%%%%%%%%%%%%%%SECTION FOUR%%%%%%%%%%%%%%%%%%%%%%%%%%%%%%%%%%%%%%%%%%%
%%%%%%%%%%%%%%%%%%%%%%%%%%%%%%%%%%%%%%%%%%%%%%%%%%%%%%%%%%%%%%%%%%%%%%%%%%%%
%%%%%%%%%%%%%%%%%%%%%%%%%%%%%%%%%%%%%%%%%%%%%%%%%%%%%%%%%%%%%%%%%%%%%%%%%%%%

\section{Dependence of detailed balance violation on coupling capacitance}
\label{sec:summary}

\begin{figure*}[t]
\includegraphics[width=2.1\columnwidth]{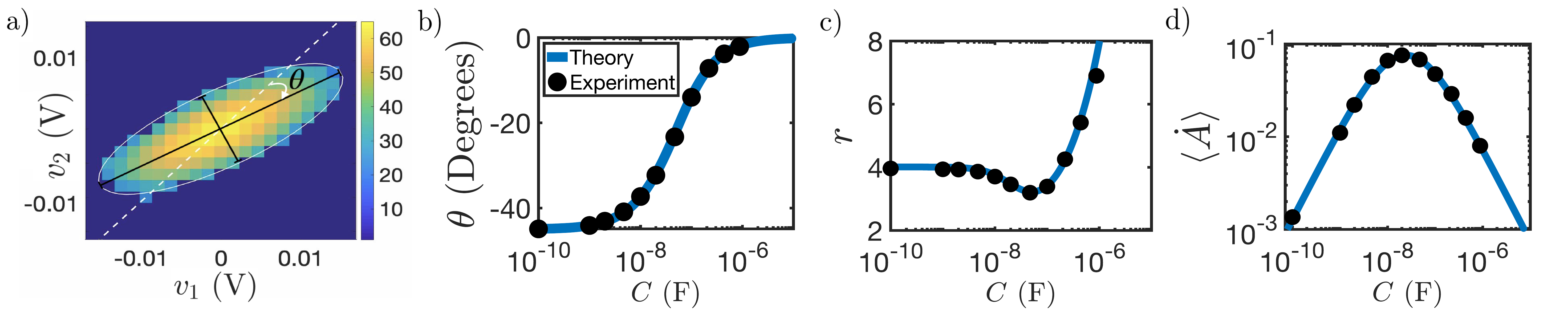}
\caption{(a) Experimentally determined elliptical probability distribution for $s_{1}^2 = 8.53\times 10^{-10}$ V$^{2}\cdot$s and  $s_2 = 5.30\times 10^{-11}$ V$^{2}$s, depicting tilt angle definition and aspect ratio determination.  (b) Tilt angle of probability ellipse vs. coupling capacitance.  (c) Probability ellipse aspect ratio vs. coupling capacitance. (d) Dimensionless stochastic area vs. coupling capacitance.}

\label{fig:Cdependence}

\end{figure*}

The central object of fluctuation statistics in the $v_1 - v_2$ plane is the second moment tensor $M$.  In (\ref{eq:ProbDensity}), the theory predicts that the stationary probability density is a Gaussian whose level curves in the  $v_1 - v_2$ plane are level curves of the stochastic action quadratic form of $M^{-1}$.  The \textit{experimental} determination of the second moment tensor consists of direct computation of averages $\langle v_i v_j \rangle$ from the recorded voltage time series.  Given the experimental second moment tensor, we construct the level curve ellipse which contains $98 \%$ of the sample points of the original voltage time series.  Figure \ref{fig:Cdependence}(a) depicts a typical example in which the probability density histogram is nicely framed by the ``$98 \%$ ellipse."  A similar bounding ellipse is superpositioned on the histogram of Fig. \ref{fig:CircuitDiagram}(a).  Such graphics verify the elliptical shape of probability density level curves, but are the orientations and shapes of the ellipses consistent with predictions according to theory?

Given the second moment tensor in the form
\begin{equation}
\label{eq:Mtensor}
M = \begin{pmatrix} \bar{m} + \delta & \mu \\ \mu & \bar{m} - \delta \end{pmatrix},
\end{equation}
the \textit{tilt angle} $\theta$, i.e., the angle between the ellipse major axis and the line $v_1 = v_2$, is given by
\begin{equation}
\label{eq:tiltangle}
\tan{\theta} = \frac{\sqrt{\mu^2 + \delta^2} - \mu - \delta}{\sqrt{\mu^2 + \delta^2} + \mu - \delta}.
\end{equation}
The \textit{aspect ratio} $r$ of a level curve ellipse, the ratio of major to minor axrs, is given by
\begin{equation}
\label{eq:aspectratio}
r^2 = \frac{\bar{m} + \sqrt{\mu^2 + \delta^2}}{\bar{m} - \sqrt{\mu^2 + \delta^2}}.
\end{equation}
Theoretical predictions of parameter dependence for tilt angle and aspect ratio are expressed in terms of the second moment tensor.  The theoretical prediction of second moment tensor according to the fluctuation-dissipation relation leads to $M$ as in (\ref{eq:Mtensor}), with
\begin{equation}
\label{eq:mu_defn}
\mu = \frac{\gamma}{2(1 + 2\gamma)} \frac{s_1^2 + s_2^2}{Rc},
\end{equation}
\begin{equation}
\label{eq:mbar_defn}
\bar{m} = \frac{ 1 + \gamma}{\gamma} \mu,
\end{equation}
\begin{equation}
\label{eq:delta_defn}
\delta = \frac{1}{2} \frac{1}{1 + \gamma} \frac{s_1^2 - s_2^2}{Rc},
\end{equation}
where $\gamma$ denotes the capacitance ratio, 

\begin{equation}
\label{eq:gamma_defn}
\gamma := \frac{C}{c}.
\end{equation}

Substituting (\ref{eq:mu_defn}) - (\ref{eq:gamma_defn}) for $\mu$, $\bar{m}$, $\delta$, and $\gamma$, respectively, into (\ref{eq:tiltangle}) and (\ref{eq:aspectratio}), we obtain the tilt angle and aspect ratio as functions of the circuit parameters and noise amplitudes.  Here, we focus on their dependences upon the coupling capacitance $C$ with all the other parameters fixed:  The fixed resistances $R_1$ and $R_2$ and the capacitors $c_1$ and $c_2$ have the same values as in preceeding sections, and the noise amplitudes are $s_1^2 = 8.53 \times 10^{-10}$ $\mathrm{V^2 s}$ and $s_2^2 = 5.30 \times 10^{-11}$ $\mathrm{V^2 s}$.  The curves in Figs. \ref{fig:Cdependence}(b), (c) are the predicted graphs of tilt angle and aspect ratio as functions of $C$.  In the limit $\frac{C}{c} \rightarrow 0$, the two $RC$ circuit elements are decoupled, and the ellipse axes are parallel to the coordinate axes.  Since the noise acting on $RC$ element 1 is stronger, the major axis aligns with the $v_1$ axis, so $\theta \rightarrow -\frac{\pi}{4}$ as $\frac{C}{c} \rightarrow 0$.  A large coupling capacitance forces the voltage difference $v_2 - v_1$ to be small, in which case we have a narrow ellipse aligned with the line $v_1 = v_2$.  Hence $\theta \rightarrow 0$ and $r \rightarrow 0$ as
$\frac{C}{c} \rightarrow \infty$.  The solid black circles in Figs. \ref{fig:Cdependence}(b) and (c) mark experimental determinations of tilt angle and aspect ratio
from long running voltage time series, one for each coupling capacitance in a sequence ranging from $C = 100 \: pF$ to $C = 880 \: nF$.  The measurement errors fall within the size of the data points.

The time rate of change of stochastic area is specified by the second moment tensor according to (\ref{eq:Mtensor}) and (\ref{eq:Adot}).  This leads to its theoretical
dependence upon circuit parameters and noise amplitudes,
\begin{equation}
\label{eq:Adot2DCircuit}
\langle \dot{A} \rangle = \frac{1}{2} \frac{\gamma}{(2 \gamma + 1)( \gamma + 1)} \frac{s_1^2 - s_2^2}{(Rc)^2}.
\end{equation}
This may also be written in non-dimensional form by measuring $\langle \dot{A} \rangle$ in units of $\frac{(s_1^2+s_2^2)}{Rc}$, resulting in
 \begin{equation}
        \langle \dot{A} \rangle = \frac{\gamma}{(2\gamma +1)(\gamma + 1)}\frac{s_1^2 - s^2_2}{s_1^2 + s^2_2}.
        \label{eq:AT_dimensionless}
    \end{equation}
From (\ref{eq:AT_dimensionless}) we see that equality of noise amplitudes, $s_1^2 = s_2^2$, implies $\langle \dot{A} \rangle = 0$, which in turn implies that the probability current is identically zero.  Figure \ref{fig:Cdependence}(d) compares the theoretical and experimental dependences of non-dimensional $\langle \dot{A} \rangle$
with coupling capacitance.  The dimensionless growth of stochastic area achieves its maximum for $\gamma = \frac{C}{c} = \frac{1}{\sqrt{2}}$.  The black dots represent experimentally determined values of $\langle \dot{A} \rangle$ in units of $\frac{(s_1^2+s_2^2)}{Rc}$ and agree closely to the theoretical prediction.  Physically, this value of coupling capacitance $\gamma$ is interpreted to provide the maximum rate at which stochastic area is generated by the system for a given difference in applied noise intensity.  Equivalently, one can say that the rate at which fluctuation loops (described in Sec. V) are swept out by the system is maximized.  On the other hand, when one examines the $\gamma$-dependence of the energy transfer rate (calculated in Sec. VI) one finds a monotonic dependence with no local maximum.  In general, the extremal behavior observed for stochastic area does not necessarily apply to other metrics that also characterize detailed balance violation in this system.

%%%%%%%%%%%%%%%%%%%%%%%%%%%%%%%%%%%%%%%%%%%%%%%%%%%%%%%%%%%%%%%%%%%%%%%%%%%%
%%%%%%%%%%%%%%%%%%%%%%%%%%%%%%%%%%%%%%%%%%%%%%%%%%%%%%%%%%%%%%%%%%%%%%%%%%%%
%%%%%%%%%%%%%%%%%%%%%%%%%%%%%%%%%%%%%%%%%%%%%%%%%%%%%%%%%%%%%%%%%%%%%%%%%%%%
%%%%%%%%%%%%%%%%%%%%%SECTION FIVE%%%%%%%%%%%%%%%%%%%%%%%%%%%%%%%%%%%%%%%%%%%
%%%%%%%%%%%%%%%%%%%%%%%%%%%%%%%%%%%%%%%%%%%%%%%%%%%%%%%%%%%%%%%%%%%%%%%%%%%%
%%%%%%%%%%%%%%%%%%%%%%%%%%%%%%%%%%%%%%%%%%%%%%%%%%%%%%%%%%%%%%%%%%%%%%%%%%%%
\section{Direct Observation of Experimental Fluctuation Loops}
\label{sec:area}
The notion of a \textit{fluctuation loop} arises from the large deviation theory of stochastic dynamical systems  \cite{Ghanta2017, Neu_Geometry_2018, WentzellFreidlin_2012, Heymann_2015, Luchinsky_RPP_1998}.  Consider trajectories in the basin of a stable critical point $\mathbf{a}$.  A displacement from $\mathbf{a}$ to a \textit{destination point} $\mathbf{b}$ - also assumed to lie in the basin of $\mathbf{a}$ - is a \textit{large deviation} if its magnitude is much larger than the standard deviation from $\mathbf{a}$.  These large deviations are rare, but when they do occur, they very nearly follow a well defined most probable \textit{fluctuation} path from $\mathbf{a}$ to $\mathbf{b}$.  After arrival in a small neighborhood of $\mathbf{b}$, the most probable continuation of the trajectory follows the deterministic \textit{relaxation} path back to $\mathbf{a}$.  If the stochastic dynamics violates detailed balance, then the fluctuation segment is \textit{not} the time reversal of the relaxation segment.  Additionally, the union of fluctuation and relaxation segments forms a closed loop containing both $\mathbf{a}$ and $\mathbf{b}$, and enclosing some nonzero area \cite{Neu_Geometry_2018}.  

Previous related observations that discern the differences between fluctuation and relaxation segments in experimental \textit{nonlinear} systems have been reported for driven micromechanical oscillators \cite{Chan_PRL_2007, Chan_PRE_2008} and also in analog electronic circuit systems \cite{Luchinsky_RPP_1998, Luchinsky_1997}.  Such experiments rely essentially on the collection of time series of sufficiently long trajectories that reach a small neighborhood $\delta R$ of a remote destination point multiple times.  Then the fluctuation segment is obtained by averaging over back histories prior to entering $\delta R$, and the relaxation path is obtained by averaging over forward histories after entering $\delta R$.

\begin{figure}[h]
\includegraphics[width=1.0\columnwidth]{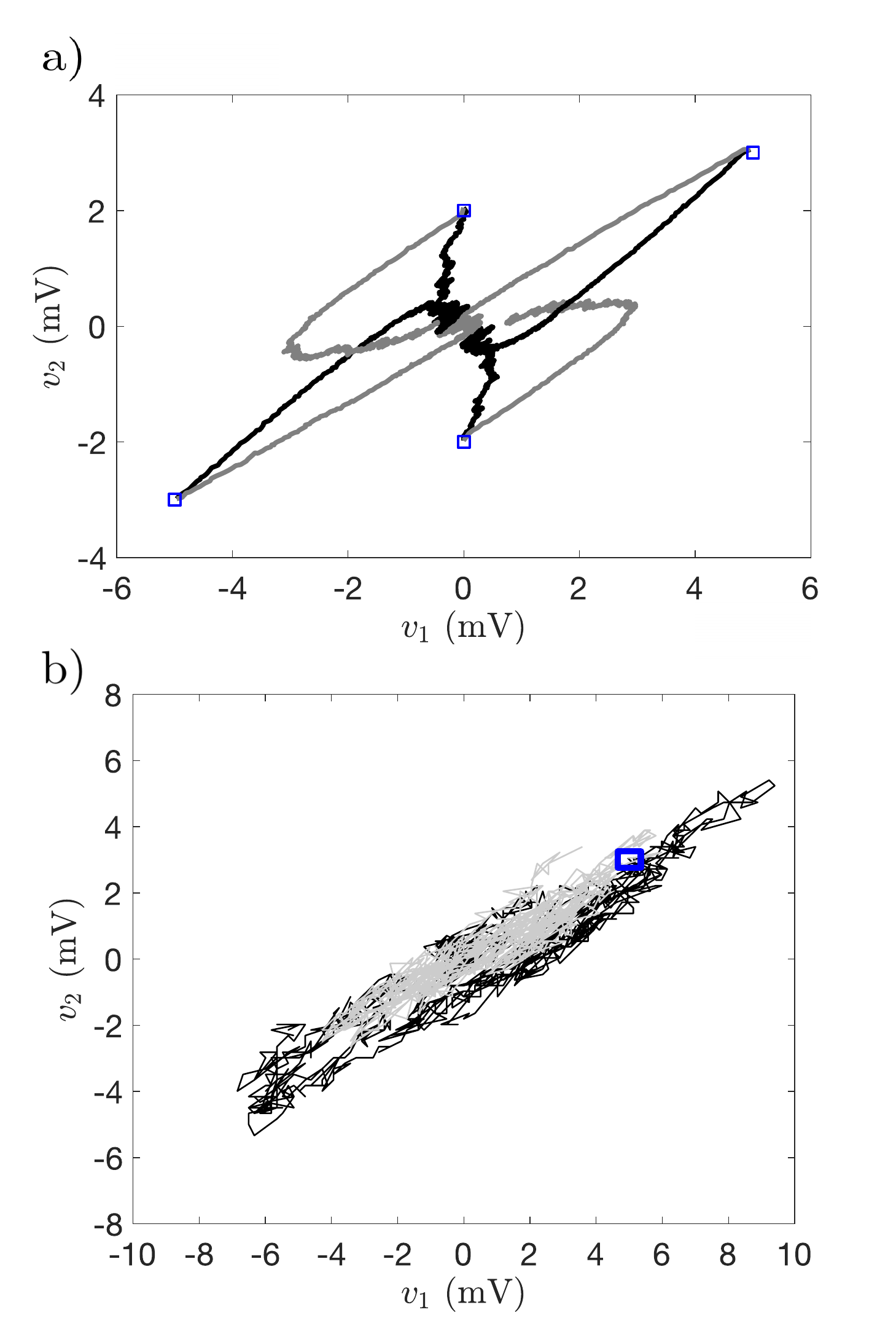}
\caption{(a) Most probable fluctuation paths (gray trajectories) and relaxation paths (black curves) determined by averaging individual experimental trajectories that pass through the indicated four target boxes.  Each curve is the result of averaging over 1000 individual trajectories that are conditioned by going forward and backward in time by 7 $Rc$ time units.  (b) One typical individual trajectory passing through the upper right target box and plotted backwards and forwards in time by 7 $Rc$ time units.}
\label{fig:FourLoops}
\end{figure}

This program is straightforward to implement for the long voltage time series recorded in our circuit experiment.  Figure \ref{fig:FourLoops}(a) shows four experimentally constructed fluctuation loops.  The circuit parameters and noise amplitudes are the same as for the probability density and current histograms in Fig. \ref{fig:CircuitDiagram}(c). The structure of the most probable fluctuation loop  The non-overlap of fluctuation and relaxation segments in Fig. \ref{fig:FourLoops}(a) indicates detailed balance violation, demonstrating that the construction of fluctuation loops is another diagnostic tool.  The required data processing is comparable to constructing the probability current in a single pixel; this follows since both constructions involve averaging over trajectories that enter a given pixel, i.e., a destination box (these are indicated in the figure). The fluctuation segments are obtained by averaging $1000$ back histories for seven $Rc$ relaxation times, and the relaxation segments, by averaging $1000$ forward histories, also for seven relaxation times.   Since orientations of fluctuation segments is outward from the origin, and the relaxation segments, inward, the sense of circulation about the loops is clockwise.  This is consistent with the clockwise circulation of probability current in Fig. \ref{fig:CircuitDiagram}(c).  In contrast to the smooth fluctuation loops of Fig. \ref{fig:FourLoops}(a), Fig. \ref{fig:FourLoops}(b) shows a single trajectory segment conditioned such that it passes through the upper righthand target box.  While one can see that this individual trajectory tends to fluctuate out from the stable equilibrium before hitting the target box, and to subsequently fall back towards the equilibrium after reaching the target box, the structure of the fluctuation loop is completely obscured by noise.

It is interesting to note the striking differences between the geometric structure of the measured fluctuation loops versus the flow lines of measured probability current.  Furthermore, it should be possible to construct the dynamical tensor $L$ and diffusion tensor $D$ from fluctuation loop measurements (in the subspace spanned by measured dynamical variables).  This might provide useful new information, for example, in experimental systems where $L$ and $D$ are not known \textit{a priori}.

We remark that the destination boxes in Fig. \ref{fig:FourLoops} represent displacements from the origin between one and ten millivolts.  As such, they are comparable to the voltage variances evident in the histogram of Fig. \ref{fig:CircuitDiagram}(c).  The loops in Fig. \ref{fig:FourLoops}(a) are not strictly speaking, ``large deviations."  Nevertheless, the averaging process resolves them with striking clarity and the measured loops agree closely with predictions of large deviation theory \cite{Ghanta2017}.  This is remarkable since predictions based on large deviation theory are expected to be strictly valid only in the small noise limit, whereas the experimental noise levels here are quite large.  These measurements thus suggest that fluctuation loops derived from large deviation theory are relatively robust and likely observable in a wide array of experimental noise-driven systems.

%%%%%%%%%%%%%%%%%%%%%%%%%%%%%%%%%%%%%%%%%%%%%%%%%%%%%%%%%%%%%%%%%%%%%%%%%%%%
%%%%%%%%%%%%%%%%%%%%%%%%%%%%%%%%%%%%%%%%%%%%%%%%%%%%%%%%%%%%%%%%%%%%%%%%%%%%
%%%%%%%%%%%%%%%%%%%%%%%%%%%%%%%%%%%%%%%%%%%%%%%%%%%%%%%%%%%%%%%%%%%%%%%%%%%%
%%%%%%%%%%%%%%%%%%%%%SECTION SIX%%%%%%%%%%%%%%%%%%%%%%%%%%%%%%%%%%%%%%%%%%%
%%%%%%%%%%%%%%%%%%%%%%%%%%%%%%%%%%%%%%%%%%%%%%%%%%%%%%%%%%%%%%%%%%%%%%%%%%%%
%%%%%%%%%%%%%%%%%%%%%%%%%%%%%%%%%%%%%%%%%%%%%%%%%%%%%%%%%%%%%%%%%%%%%%%%%%%%

\section{Relationship of stochastic area to other metrics for characterizing nonequilibrium systems}
\label{sec:Disc}

The literature presents several different metrics for characterizing nonequilibrium dynamics.  Among these, the \textit{cycling frequency} or \textit{phase space torque} \cite{Weiss_PRE_2007, Gnesotto_2018, Mura_2018}, like stochastic area, are directly related to the geometry of the stochastic dynamical system and its trajectories in phase space.  Another set of metrics are related to the physical processes of \textit{energy transfer} and \textit{entropy production} \cite{Ciliberto_PRL_2013, Ciliberto_JSM_2013b}.  In this section, we discuss the relationships between these different metrics in the context of our experimental circuit system. 

Recently, Mura \textit{et al.} \cite{Mura_2018} have shown that any two state variables
of a large linear network satisfy a reduced two-dimensional stochastic dynamics, and the probability current in the plane of any two variables is given by
\begin{equation}
\label{eq:BProbCurrent}
\mathbf{j} = B \mathbf{v} \rho.
\end{equation}
Here, $\mathbf{v} = \begin{pmatrix} v_1 \\ v_2 \end{pmatrix}$ is the two-vector of state variables, $\rho$ is the stationary probability density of $\mathbf{v}$, and the tensor $B$ has pure imaginary eigenvalues $i \omega,  -i \omega$.  The real number $\omega$ is called the\textit{ cycling frequency}.  Comparing (\ref{eq:BProbCurrent}) with (\ref{eq:StaProbCurrent}), we make the identification
\begin{equation}
B = \begin{pmatrix} 0 & -\Omega \\ \Omega & 0 \end{pmatrix} M^{-1} = \begin{pmatrix} 0 & -\Omega \\ \Omega & 0 \end{pmatrix}  \begin{pmatrix} \bar{m} + \delta & \mu \\ \mu  & \bar{m} - \delta \end{pmatrix} ^{-1}.
\end{equation}
The eigenvalues are readily computed to be $i \omega,  -i \omega$, with
\begin{equation}
\omega = \frac{\Omega}{\sqrt{\bar{m}^2 -\delta^2 - \mu^2}} = \frac{\Omega}{\sqrt{\det M}},
\end{equation}
and we see that the cycling frequency $\omega$ can be expressed as the rate of change $\Omega$ of stochastic area (cf. (\ref{eq:Adot}) above) divided by $\sqrt{\det M}$.  

It is also possible to demonstrate the proportionality of the cycling frequency to the \textit{phase space torque} \cite{Mura_2018}.  For two-dimensional stochastic dynamics (e.g., the circuit model studied in this paper (\ref{eq:CircuitModel})),
this torque is defined to be
\begin{equation}
\tau := \langle v_2 (L\mathbf{v})_1 - v_1 (L\mathbf{v})_2 \rangle,
\end{equation}
where $L\mathbf{v}$ can be viewed as the ``deterministic force" acting in the $v_1, v_2$ plane.  We calculate
\begin{eqnarray}
\label{eq:2Dtorque}
\tau_{21} &:=&  L_{1j} \langle v_j v_2 \rangle -  L_{2j} \langle v_j v_1 \rangle \\ \nonumber  &=& (LM)_{12} - (LM)_{21} \\ \nonumber &=& (LM + D)_{12} - (LM + D)_{21} = 2\Omega.
\end{eqnarray}
Hence, the phase space torque is twice the rate of change of stochastic area.

For linear stochastic dynamics with dimension $N$ greater than two, the phase space torque $\tau$ is an $N \times N$ antisymmetric tensor \cite{Mura_2018}, and it is straightforward to show that the generalization of (\ref{eq:2Dtorque}) can be written as the tensorial relation
\begin{equation}
\tau = 2(LM + D).
\end{equation}
In Ghanta \textit{et al.} \cite{Ghanta2017}, it was shown that $-(LM + D)$ is the long time asymptotic time rate of change of the \textit{stochastic area tensor}
\begin{equation}
A(t) := \int_0^t \langle (\mathbf{v} \dot{\mathbf{v}}^T - \dot{\mathbf{v}} \mathbf{v}^T)(t') \rangle dt'.
\end{equation}

The notion of cycling frequency has its natural generalization to $N$ dimensions as well.  The current density on $N$-dimensional phase space is given by (\ref{eq:BProbCurrent}) with $\mathbf{v}$ now interpreted as the state vector in $\mathbb{R}^N$, and
\begin{equation}
B := (LM + D)M^{-1},
\end{equation}
where the identification of $B$ in terms of $L$, $M$, and $D$ follows from (\ref{eq:StaProbCurrent}).  We now show that the eigenvalues of $B$ are pure imaginary.  We can reformulate the eigenvalue problem for $B$, i.e., $B \mathbf{v} = \lambda \mathbf{v}$, as
\begin{equation}
\label{eq:Eigenv}
(LM + D) \mathbf{v} = \lambda M \mathbf{v}.
\end{equation}
Assuming that the symmetric second moment tensor $M$ is nonsingular, it has a nonsingular square root.  We introduce 
\begin{equation}
\mathbf{y} := M^{\frac{1}{2}} \mathbf{v}
\end{equation}
in place of $\mathbf{v}$, and reformulate (\ref{eq:Eigenv}) as 
\begin{equation}
\label{eq:Eigeny}
B' \mathbf{y} = \lambda \mathbf{y}, \; B' := M^{-\frac{1}{2}} (LM + D) M^{-\frac{1}{2}}.
\end{equation}
The antisymmetry of $LM + D$ implies that $B'$ is antisymmetric.  The antisymmetry (and reality) of $B'$ implies that all eigenvalues are pure imaginary and that the \textit{non-zero} imaginary parts occur in complex conjugate pairs \cite{Horn_1994}.  In general, this gives rise to a collection of cycling frequencies for the $N$-dimensional system.  To summarize, we see that the stochastic area tensor, the cycling frequency, and the phase space torque are all related to the tensor $LM +D$ whose antisymmetry is the direct expression of the fluctuation-dissipation relation \cite{Weiss_Tellus_2003, Ghanta2017}.   

Next, we discuss how physically-based metrics of detailed balance violation such as energy transport and entropy production also have their immediate connection  to $LM + D$, hence to stochastic area, phase space torque, and cycling frequency.  In the circuit model (\ref{eq:CircuitModel}), the energy in all the capacitors is 
\begin{equation}
\label{eq:2DEnergy}
E = \frac{c}{2} (v_1^2 + v_2^2) + \frac{C}{2} (v_1 - v_2)^2.
\end{equation}
Time differentiation of (\ref{eq:2DEnergy}) and use of the stochastic ODE (\ref{eq:CircuitModel}) leads to the \textit{energy identity}
\begin{equation}
\dot{E} = p_1 + p_2,
\end{equation}
where
\begin{equation}
p_1 := \frac{v_1 s_1 w_1}{R} - \frac{v_1^2}{R},
\end{equation}
and $p_2$ is defined analogously.  We recognize $-v_1^2 / R$ as energy dissipated by the Joule heating of the resistor in $Rc$ circuit one.  It is natural to interpret $v_1 s_1 w_1 / R$ as the work done on the whole circuit by the channel one noise.  For stationary statistics, we have $\langle \dot{E} \rangle = 0$, so $\langle p_1 \rangle +
\langle p_2 \rangle = 0$.  We interpret the common value $p$ of $\langle p_1 \rangle$ and $-\langle p_2 \rangle$ as the average rate of 
energy transfer from circuit one to circuit two.

We now relate $p$ to the rate of change of stochastic area.  The one-component of the stochastic ODE (\ref{eq:CircuitModel}) may be written as
\begin{equation}
-\frac{v_1}{R} + \frac{s_1w_1}{R} = C(\dot{v}_1 - \dot{v}_2 ) + c \dot{v}_1.
\end{equation}
Multiplying by $v_1$ and taking the ensemble average, we have
\begin{equation}
\label{eq:EnergyArea}
p = -C \langle v_1 \dot{v}_2 \rangle = \frac{C}{2} \langle v_2 \dot{v}_1 - v_1 \dot{v}_2 \rangle = C \langle \dot{A} \rangle.
\end{equation}
Substituting for $\langle \dot{A} \rangle$ from (\ref{eq:Adot2DCircuit}), we have
\begin{equation}
\label{eq:PowerTransfer}
p = \frac{1}{2} \frac{\gamma^2}{(2\gamma + 1)(\gamma + 1)} \frac{1}{Rc} (\frac{s_1^2}{R} - \frac{s_2^2}{R}).
\end{equation}
This connection between stochastic area and heat transfer rates extends to network with many degrees of freedom connected to an assortment of thermal baths, all with their associated dissipation and noise. In analogy with (\ref{eq:EnergyArea}), it is the stochastic area \textit{tensor} which informs heat transfer rates between the different nodes.  This is the subject of ongoing work by the authors \cite{Neu_2018}.

Ciliberto \textit{et al.} \cite{Ciliberto_PRL_2013, Ciliberto_JSM_2013b} have studied energy and entropy transport both experimentally and theoretically for a similar coupled circuit that utilizes only intrinsice thermal noise sources of the resistances.  To make a connection with this work, we may quantify energy transport
and entropy production in terms of effective temperatures $T_1, T_2$ of the resistors in circuits one and two.  These effective temperatures are related to noise amplitudes $s_1, s_2$ according to the Nyquist formula,  $s_1 = \sqrt{R k_B T_1}, \; s_2 = \sqrt{R k_B T_2}$.  Thus, one may rewrite (\ref{eq:PowerTransfer}) for the average energy transfer rate in terms of these effective temperatures as
\begin{equation}
p = \frac{1}{2} \frac{\gamma^2}{(2\gamma + 1)(\gamma + 1)} \frac{k_B(T_1 - T_2)}{Rc}.
\end{equation}
The \textit{entropy production rate} associated with this energy transfer is
\begin{equation}
\frac{-p}{T_1} + \frac{p}{T_2} = \frac{1}{2} \frac{\gamma^2}{(2\gamma + 1)(\gamma + 1)} \frac{k_B}{Rc} \frac{(T_1 - T_2)^2}{T_1 T_2}.
\end{equation}
We note that, unlike the expression for $\gamma$ dependence of stochastic area (cf. (\ref{eq:AT_exp}) above), neither energy nor entropy production rates exhibit a local maximum as coupling capacitance is varied.  Instead, they both increase monotonically with coupling capacitance $C$, or equivalently $\gamma$. 

In this section, we have shown that several metrics for characterizing nonequilibrium dynamics are closely related to one another, at least for relatively simple noise-driven linear systems (e.g., two-dimensional coupled linear circuits) with well-understood theoretical descriptions.  For systems such as these, the choice as to which metric or combination of metrics to use for analysis of experimental data largely depends on the properties to be measured and characterized. On the other hand, for experimental systems where the underlying dynamics may be unknown or for which an energy function is not available, the use of geometric metrics (e.g., cycling frequencies, phase space torque, and stochastic area tensor) should remain feasible as these techniques rely only the capability to measure time series of (at least) two independent dynamical variables.

%%%%%%%%%%%%%%%%%%%%%%%%%%%%%%%%%%%%%%%%%%%%%%%%%%%%%%%%%%%%%%%%%%%%%%%%%%%%
%%%%%%%%%%%%%%%%%%%%%%%%%%%%%%%%%%%%%%%%%%%%%%%%%%%%%%%%%%%%%%%%%%%%%%%%%%%%
%%%%%%%%%%%%%%%%%%%%%%%%%%%%%%%%%%%%%%%%%%%%%%%%%%%%%%%%%%%%%%%%%%%%%%%%%%%%
%%%%%%%%%%%%%%%%%%%%%SECTION SEVEN%%%%%%%%%%%%%%%%%%%%%%%%%%%%%%%%%%%%%%%%%%%
%%%%%%%%%%%%%%%%%%%%%%%%%%%%%%%%%%%%%%%%%%%%%%%%%%%%%%%%%%%%%%%%%%%%%%%%%%%%
%%%%%%%%%%%%%%%%%%%%%%%%%%%%%%%%%%%%%%%%%%%%%%%%%%%%%%%%%%%%%%%%%%%%%%%%%%%%

\section{Conclusions and a historical connection}
\label{sec:summary}

In this paper, we have presented data and analysis from a real circuit experiment that shows detailed balance violation when driven by external noise generators.  A central result of this paper concerns the utility of the stochastic area as a quantitative indicator of detailed balance breaking in \textit{experimental} noise-driven linear dynamical systems.  This metric can likely be implemented for a wide range of noise-driven systems.  The application to any system requires the measurement of the rate of change of area swept out in the plane of any two independent observables.  A nonzero average rate of change indicates violation of detailed balance.  In this sense, stochastic area provides a widely applicable means for quantifying detailed balance violation.  

In Sec. IV we showed that the rate of change of stochastic area has its largest magnitude for parameter choice $\gamma =  \frac{1}{\sqrt{2}}$.  One might ask whether this parameter choice also maximizes the rate of energy transfer from one $Rc$ circuit element to the other, since  nonzero energy flow is also an indicator of detailed balance violation.  
Like the stochastic area, the energy transfer rate vanishes only if there is detailed balance. However, unlike the stochastic area, we have seen that the energy transfer and entropy production rates are both monotonically increasing with the coupling capacitance.

 The stochastic area has a compelling connection to Onsager's \textit{theoretical} characterization of thermodynamic fluctuations.  Onsager \cite{Onsager_1931} proposed that thermodynamic equilibrium upholds a certain symmetry of temporal correlations as follows:  let $x(t)$ and $y(t)$ be stationary random processes representing fluctuations of two state variables.  For \textit{equilibrium} statistics, the correlation function $\langle x(t) y(t') \rangle$ is invariant under translation of times $t$ and $t'$ by the same constant (stationary stochastic processes) and also invariant under interchange of $t$ and $t'$.  Onsager calls this exchange symmetry the \textit{principle of microscopic reversibility}.  Due to the exchange symmetry, equilibrium statistics does not betray the forward direction of time.  The connection to stochastic area is immediate:  According to microscopic reversibility, we have
\begin{equation}
\frac{1}{2 \tau} \langle x(t) y(t + \tau) - x(t + \tau) y(t) \rangle = 0.
\end{equation}
for all $t$ and $\tau$.  In the limit $\tau \rightarrow 0$, the LHS reduces to
\begin{equation}
\frac{1}{2} \langle (x \dot{y} - y \dot{x})(t) \rangle.
\end{equation}
This is none other than the ensemble-averaged time rate of change of stochastic area
\begin{equation}
A(t) = \frac{1}{2} \int_0^t (x \dot{y} - y \dot{x})(t') dt'.
\end{equation}
Hence, ensemble-averaged stochastic area has zero rate of change for equilibrium statistics.  Nonzero growth of stochastic area indicates violation of Onsager's microscopic reversibility.

We conclude by posing a related open question concerning applicability of these methods to higher dimensional systems. Experiments typically probe only a few of many independent state variables.  This is certainly the case for the experiments on active biological systems as in \cite{Battle_Science_2016, Gladrow_2016}.   Probability density histograms constructed from time series of observables are obviously projections of the probability density on the whole state space.  The formal algorithms to construct probability current histograms on the subpace of observables generally remain operable, but what do these formal probability currents really mean?  Do they really describe transport of the reduced probability density in the space of observables, or is there a mismatch which reflects the presence of ignored dimensions?

%%%%%%%%%%%%%%%%%%%%%%%%%%%%%%%%%%%%%%%%%%%%%%%%%%%%%%%%%%%%%%%%%%%%%%%%%%%%
%%%%%%%%%%%%%%%%%%%%%%%%%%%%%%%%%%%%%%%%%%%%%%%%%%%%%%%%%%%%%%%%%%%%%%%%%%%%
%%%%%%%%%%%%%%%%%%%%%%%%%%%%%%%%%%%%%%%%%%%%%%%%%%%%%%%%%%%%%%%%%%%%%%%%%%%%
%%%%%%%%%%%%%%%%%%%%%ACKNOWLEDGEMENTS%%%%%%%%%%%%%%%%%%%%%%%%%%%%%%%%%%%%%%%%%%%%%
%%%%%%%%%%%%%%%%%%%%%%%%%%%%%%%%%%%%%%%%%%%%%%%%%%%%%%%%%%%%%%%%%%%%%%%%%%%%
%%%%%%%%%%%%%%%%%%%%%%%%%%%%%%%%%%%%%%%%%%%%%%%%%%%%%%%%%%%%%%%%%%%%%%%%%%%%
\acknowledgements

The authors gratefully acknowledge helpful conversations with Gleb Finkelstein.

%%%%%%%%%%%%%%%%%%%%%%%%%%%%%%%%%%%%%%%%%%%%%%%%%%%%%%%%%%%%%%%%%%%%%%%%%%%%
%%%%%%%%%%%%%%%%%%%%%%%%%%%%%%%%%%%%%%%%%%%%%%%%%%%%%%%%%%%%%%%%%%%%%%%%%%%%
%%%%%%%%%%%%%%%%%%%%%%%%%%%%%%%%%%%%%%%%%%%%%%%%%%%%%%%%%%%%%%%%%%%%%%%%%%%%
%%%%%%%%%%%%%%%%%%%%%APPENDICES%%%%%%%%%%%%%%%%%%%%%%%%%%%%%%%%%%%%%%%%%%%%%
%%%%%%%%%%%%%%%%%%%%%%%%%%%%%%%%%%%%%%%%%%%%%%%%%%%%%%%%%%%%%%%%%%%%%%%%%%%%
%%%%%%%%%%%%%%%%%%%%%%%%%%%%%%%%%%%%%%%%%%%%%%%%%%%%%%%%%%%%%%%%%%%%%%%%%%%%
\appendix
\section{Experimental determination of effective noise amplitudes}
\label{sec:appendix1}
Recall that the filtered noise voltages $s_1(t)$ and $s_2(t)$ injected into the circuit are modeled as multiples $s_1 w_1(t)$ and $s_2 w_2(t)$ of unit white noises $w_1(t)$ and $w_2(t)$.  Here, we present the determinations of the effective noise amplitudes $s_1$ and $s_2$ from recorded time series of $s_1(t)$ and $s_2(t)$.  Letting $s(t)$ refer to a stationary stochastic process such as $s_1(t)$ or $s_2(t)$, and  assuming that $s(t)$ has zero mean, its integral
\begin{equation}
B(t) := \int_0^t s(t') dt'
\end{equation}
also has zero mean.  Its variance is
\begin{equation}
\langle B^2 (t) \rangle = \int_0^t \int_0^t c(t' - t'') dt' dt'',
\end{equation}
where $c(t)$ is the correlation function such that
\begin{equation}
c(t' - t'') = \langle s(t') s(t'') \rangle.
\end{equation}
Assuming $c(t)$ is integrable, we define the \textit{correlation time} $t^*$ as
\begin{equation}
t^* := \frac{1}{2c(0)} \int_{-\infty}^\infty c(t') dt',
\end{equation}
which characterizes the width of the main support of $c(t)$ about the origin.  In the limit $t >> t^*$, (A2) reduces asymptotically to
\begin{equation}
\langle B^2(t) \rangle \sim \Big\{ \int_{-\infty}^\infty c(t') dt' \Big\} t.
\end{equation}
Variance proportional to time is the signature feature of Brownian motion.  Hence, we say that $B(t)$ is asymptotic to a Brownian motion for $t >> t^*$.   If we replace the actual noise $s(t)$ in (A2) by the multiple $s w(t)$ of unit white noise $w(t)$, then $B(t)$ is exactly Brownian motion, with 
\begin{equation}
\langle B^2(t) \rangle = 2 s^2 t.
\end{equation}
Comparing (A5) and (A6), we identify the effective noise amplitude of $s(t)$,
\begin{equation}
s^2 = \frac{1}{2} \int_{-\infty}^\infty c(t') dt'.
\end{equation}
An alternative characterization of noise amplitude,
\begin{equation}
s^2 = \frac{1}{2} \lim_{t \rightarrow \infty} \frac{ \langle B^2(t) \rangle }{ t},
\end{equation}
follows from (A5) and (A7), and is the basis for its experimental determination.

Practical implementation starts with the recording of a long time series of $s(t)$.  The sampling interval $\tau$ should be much smaller than the noise correlation time $t^*$.  Here, we use the smallest sampling interval permitted by the multichannel analog-to-digital converter, $\tau = 2 \; \text{ns}  << t^* \approx 400$ ns.  We divide the complete time series into a large number $N >> 1$ of sub-series, each of which consists of $n >> 1$ sequential data points, such that $n \tau >> t^*$.  Indexing each of the sub-series by integer $k$, we have an ensemble of discrete analogs of the integral (A1),
\begin{equation}
B_k(t = n \tau) = \sum_{j = 0}^{n-1} s(j \tau + kn \tau) \tau,
\end{equation}
for $k = 1, ... N$.  For each sub-series, we calculate the mean square displacement normalized by elapsed time $n \tau$, 
\begin{equation}
\frac{1}{2} \frac{B_k^2(n \tau)}{n \tau}.
\end{equation}
The average of these values over all sub-series provides an experimental determination of the effective noise amplitude $s^2$.

\section{The discrete stochastic area formula}
\label{sec:appendix2}

For what range of sampling intervals is the ensemble-averaged finite difference $\langle (DA)(t) \rangle$ in (\ref{eq:DA_exp}) a good approximation to the corresponding theoretical expression $\langle \dot{A} \rangle$, cf. (\ref{eq:Adot})?  One concern is that the voltage time series with sampling time larger than the noise correlation time, $\tau >> t^* \approx 400$ ns, does not detect the short time fluctuations between successive sample points.  Does this matter?  A simple analysis settles this question.  For any realization of the noise vector $\mathbf{w}(t)$ in (4), the corresponding trajectory in the stationary ensemble is
\begin{equation}
\mathbf{v}(t) = \int_0^\infty e^{Lt'} \sigma \mathbf{w}(t - t') dt'.
\end{equation}
We calculate
\begin{eqnarray}
& & \langle \mathbf{v}(t) \mathbf{v}(t + \tau)^T \rangle =  \nonumber \\
& & \int_0^\infty \int_0^\infty e^{Lt'} \sigma \langle \mathbf{w}(t - t') \mathbf{w} (t + \tau - t'')^T \rangle \sigma^T e^{L^T t''} dt' dt'' = \nonumber \\
& & \int_0^\infty \int_0^\infty e^{Lt'} \sigma \delta ( \tau - t'' + t') \sigma^T e^{L^T t''} dt' dt'' = \nonumber \\
& & \int_0^\infty  e^{Lt'} \sigma  \sigma^T e^{L^T (t' + \tau)} dt'  = \nonumber \\
& & (\int_0^\infty  e^{Lt'} D e^{L^T t'}  dt' ) e^{L^T \tau} = Me^{L^T \tau}.
\end{eqnarray}
Similarly,
\begin{equation}
\langle \mathbf{v}(t + \tau) \mathbf{v}(t)^T \rangle = e^{L \tau} M.
\end{equation}
Hence,
\begin{equation}
\langle (DA)(t) \rangle = \frac{1}{2} (M e^{L^T \tau} - e^{L \tau} M)_{12}.
\end{equation}
If the sampling time $\tau$ is much shorter than the relaxation time $Rc$ associated with the dynmical matrix $L$, then (B4) asymptotically reduces to
\begin{equation}
\langle (DA)(t) \rangle  \sim \frac{1}{2} (ML^T - LM)_{12} = -(LM + D)_{12} = \Omega.
\end{equation}
Thus, we see that it is the relaxation time $Rc$ and not the much shorter noise correlation time $t^*$ which sets the upper bound on the sampling interval.

\end{document}